\pdfoutput=1

\documentclass[showpacs, pre]{revtex4}   

\usepackage{amsmath}
\usepackage{amssymb}
\usepackage{times}
\usepackage{verbatim}
\usepackage[applemac]{inputenc}
\usepackage[T1]{fontenc}
\usepackage{lmodern}
\usepackage{wrapfig}
\usepackage{graphicx}
\usepackage{rotating}
\usepackage{multirow}
\usepackage{float}

\DeclareGraphicsExtensions{.pdf}

\begin{document}

\title{Opportunistic Migration in Spatial Evolutionary Games}

\author{Pierre Buesser}
\email{pierre.buesser@unil.ch}
\affiliation{Information Systems Institute, HEC, University of Lausanne, Switzerland}

\author{Marco Tomassini}
\email{marco.tomassini@unil.ch}
\affiliation{Information Systems Institute, HEC, University of Lausanne, Switzerland}

\author{Alberto Antonioni}
\email{marco.tomassini@unil.ch}
\affiliation{Information Systems Institute, HEC, University of Lausanne, Switzerland}


\begin{abstract}

We study evolutionary games in a spatial diluted grid environment in which agents strategically interact locally but can also
opportunistically move to other positions within a given migration radius. Using the imitation of the best rule for strategy revision,
it is shown that cooperation may evolve and be stable in the Prisoner's Dilemma game space for several migration distances
but only for small game interaction radius while the Stag Hunt class of games become fully cooperative. We also show that 
only a few trials are needed for cooperation to evolve, i.e. searching costs are not an issue. When the stochastic Fermi strategy
update protocol is used cooperation cannot evolve in the Prisoner's Dilemma if the selection intensity is high in spite of
opportunistic migration. However, when
imitation becomes more random, fully or partially cooperative states are reached in all games for all migration distances tested
and for short to intermediate interaction radii.

\end{abstract}

%
%

\maketitle



\section{Introduction}
\label{intro}

Spatially embedded systems are very important in biological and social sciences since most interactions
among living beings or artificial actors take place in physical two- or three-dimensional space~\cite{Geometry}.
Along these lines, game-theoretical interactions among spatially embedded
agents distributed according to a fixed structure in the plane have been studied in detail, starting from the pioneering works of
Axelrod~\cite{axe84} and Nowak and May~\cite{nowakmay92}. The related literature is very large; see, for instance, the
review article by Nowak and Sigmund~\cite{nowak-sig-00} and references therein for a synthesis. Most of this work was
based on populations of agents arranged according to planar regular grids for mathematical simplicity and ease of
numerical simulation. Recently, some extensions to more general spatial networks 
have been discussed in~\cite{buesser-tom-space}. The study of strategic behavior on fixed spatial structures is 
necessary in order to 
understand the basic mechanisms that may lead to socially efficient global outcomes such as cooperation and
coordination. However, in the majority of real situations both in biology and in human societies, actors have the
possibility to move around in space.
Many examples can be found in biological and ecological sciences, in human populations, and in engineered systems
such as ad hoc networks of mobile communicating devices or robot teams.
Mobility may have positive or negative effects on cooperation, depending on several factors. An early investigation was
carried out by
Enquist and Leimar~\cite{Enquist} who
concluded that mobility may seriously restrict the
evolution of cooperation.  In the last decade there have
been several new studies of the influence of mobility on the behavior of various games in spatial
environments representing essentially two strands of research: one in which the movement of agents is seen as a random
walk, and a second one in which movement may contain random elements but it is purposeful, or strategy-driven. 

Random diffusion of mobile agents through space, either in
 continuous space or, more commonly,  on diluted grids has been investigated in~\cite{Meloni,Arenzon1,Arenzon2}. 
In the present study we focus on situations where, instead of randomly diffusing, agents possess some basic cognitive
abilities and they actively seek
to improve their situation by moving in space represented as a discrete grid in which part of the available
sites are empty and can thus be the target of the displacement. This approach has been followed, for example,
in~\cite{helbingPNAS,Helb-Mobil,adaptive-mig,droz-09,chen-perc,reput-migr,SwarmPD,aktipis}. The mechanisms invoked range from 
success-driven migration~\cite{helbingPNAS}, adaptive migration~\cite{adaptive-mig}, reputation-based
 migration~\cite{reput-migr}, risk-based migration~\cite{chen-perc}, flocking behavior~\cite{SwarmPD}, and cooperators walking away from defectors~\cite{aktipis}.
In spite of the difference among the proposed models, the
general qualitative message of this work is that purposeful contingent movement may lead to highly cooperating stable or quasi-stable
 population states if some conditions are satisfied. 

 Our approach is based on numerical simulation and is inspired by the work of Helbing and Yu~\cite{Helb-Mobil,helbingPNAS} which they call
 ``success-driven migration'' and which has been shown to be able to produce highly cooperative states. In this model, locally interacting agents playing either defection or cooperation
 in a two-person Prisoner's Dilemma are initially  randomly distributed on a grid such that there are empty grid points. Agents update their strategies according to their own payoff and the payoff earned by their first neighbours  but they can also ``explore'' an extended square neighborhood by
 testing all the empty positions up to a given distance. If the player finds that it would be more profitable to move
 to one of these positions then she does it, choosing the best one among those tested, otherwise she stays at her current place. Helbing and Yu find that robust cooperation states
 may be reached by this mechanism, even in the presence of random noise in the form of random strategy mutations
 and random agent relocation.  Our study builds upon this work in several ways. In the first place, whilst Helbing and Yu
 had a single game neighborhood and migration neighborhood, we systematically investigate these
 two parameters showing that only some combination do foster cooperation using success-driven migration. Secondly,
 cost issues are not taken into account in~\cite{helbingPNAS}. However, it is clear that moving around to test the ground
 is a costly activity. In a biological setting, this could mean using up energy coming from metabolic activity, and this
 energy could be in short supply. In a human society setting, it is the search time that could be limited in a way or another. Additionally to physical energy, cognitive abilities could also limit the search. 
 We present results for a whole game phase
 space including the Hawk-Dove class of games, and the Stag Hunt coordination class. Helbing's and Yu's agents
 based their strategy change on the imitation of the most successful neighbour in terms of accumulated payoff. We kept
 this rule but also added the Fermi strategy-updating rule, a choice that allows us to introduce a parametrized
amount of imitation noise. With the imitiation of the best policy we find that cooperation prevails in the Stag Hunt and may evolve
in the Prisoner's Dilemma for small interaction radius. With the Fermi rule fully cooperative states are reached for the standard 
neighborhoods independently of the migration distances when the rate of random strategy imitation is high enough.

\section{Methods}

\subsection{The Games Studied}

We investigate three classical two-person, two-strategy, symmetric games classes, namely the Prisoner's
Dilemma (PD),
the Hawk-Dove Game (HD), and the Stag Hunt (SH). These three games are simple metaphors for different kinds
of dilemmas that arise when individual and social interests collide. The Harmony game (H) is included for completeness but it
doesn't originate any conflict. 
The main features of these games are well known; more detailed accounts
can be found elsewhere e.g.~\cite{Hofbauer1998,vega-redondo-03,weibull95}.
The games have the generic payoff matrix $M$ (equation~\ref{eq:payoff}) which refers to the payoffs of the row player. The payoff matrix for the column player
is simply the transpose $M^\top$ since the game is symmetric.

\vspace{-0.5cm}
\begin{equation}
	\bordermatrix{\text{}&C& D\cr
	C&R&S\cr
	D&T&P\cr
	}
\label{eq:payoff}
\end{equation}
\vspace{0.05cm}
\noindent The set of strategies is $\Lambda=\{C,D\}$, where $C$ stands for ``cooperation'' and $D$ means ``defection''.
In the payoff matrix $R$ stands for the \textit{reward}
the two players receive if they
both cooperate, $P$ is the \textit{punishment} if they both defect, and $T$  is the
\textit{temptation}, i.e.~the payoff that a player receives if he defects while the
other cooperates getting the \textit{sucker's payoff} $S$.
For the PD, the payoff values are ordered such that $T > R > P > S$. 
Defection is always the best rational individual choice, so that 
$(D,D)$ is the unique Nash Equilibrium (NE).
In the HD game the payoff ordering is $T > R > S > P$. Thus,
when both players defect they each get the lowest payoff. 
$(C,D)$ and $(D,C)$ are NE of the game in pure strategies. There is
a third equilibrium in mixed strategies which is the only dynamically stable equilibrium~\cite{weibull95,Hofbauer1998}.
In the SH game, the ordering is $R > T > P > S$, which means that mutual cooperation $(C,C)$ is the best outcome and a NE.   The second NE, where both players defect
is less efficient but also less risky.  The third NE is in mixed strategies but it is evolutionarily unstable~\cite{weibull95,Hofbauer1998}.
Finally, in the H game $R>S>T>P$ or $R>T>S>P $. In this case $C$ strongly dominates $D$ and
the trivial unique NE is $(C,C)$. The game is non-conflictual by definition; it is mentioned to complete the quadrants of the parameter space.

\begin{figure*}[ht!]
\begin{center}
\begin{tabular} {cccccccc} 
 \includegraphics[width=6cm]{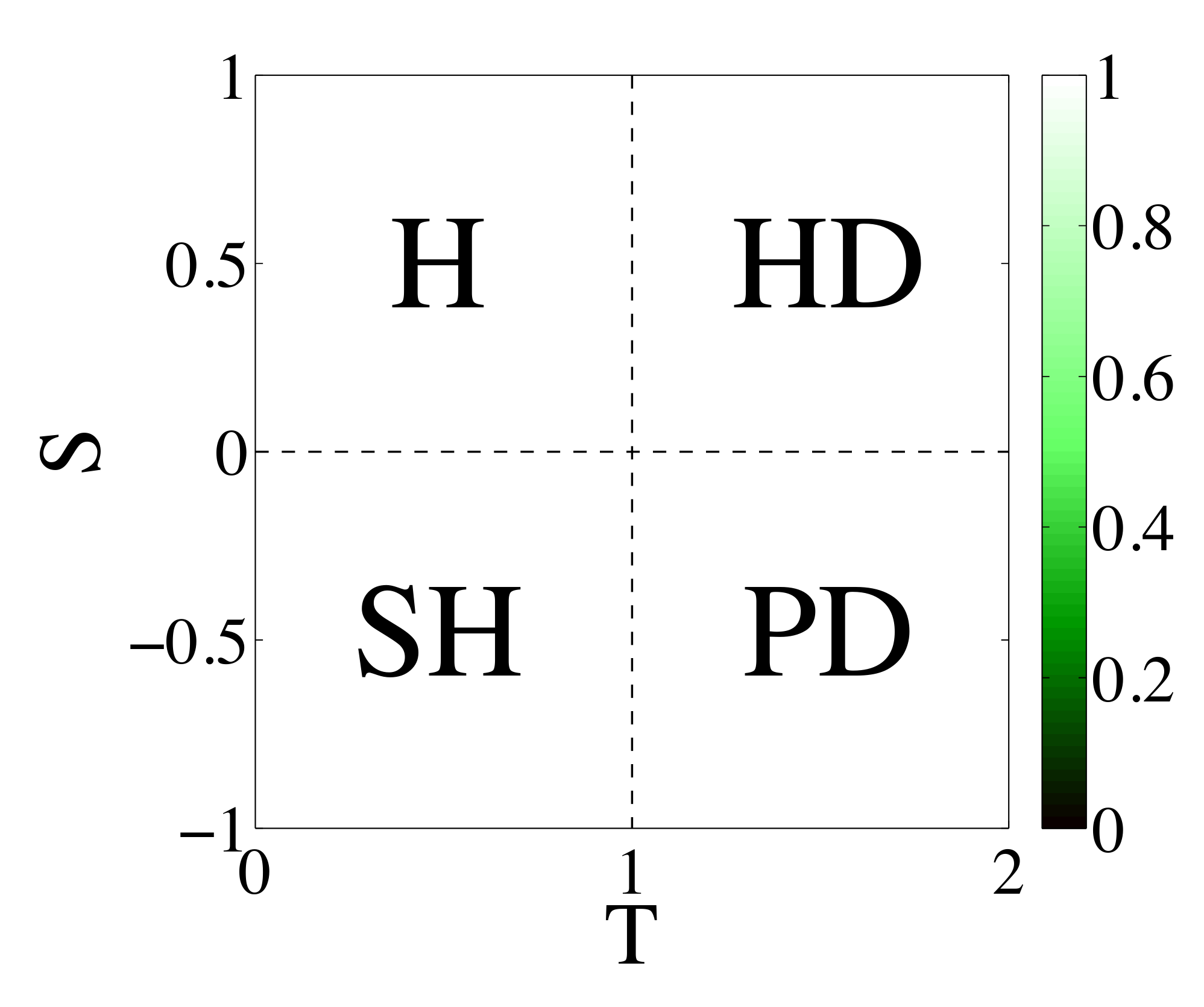} &
   \includegraphics[width=6cm]{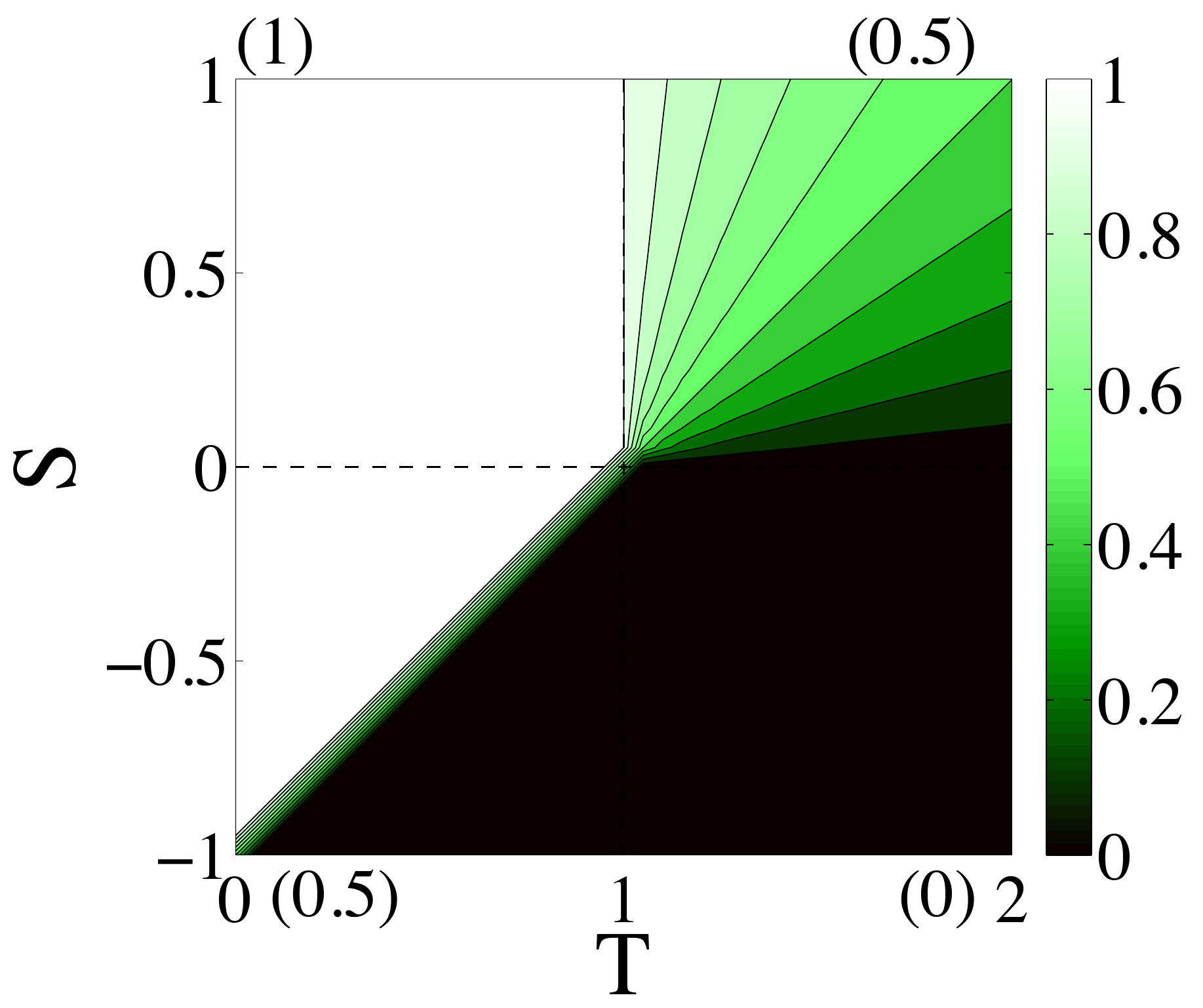} &
\end{tabular}
\caption{(Color online)  Left image: The games phase space (H= Harmony, HD = Hawk-Dove, PD = Prisoner's Dilemma, and SH = Stag Hunt) as a function of $S,T$ ($R=1$, $P=0$). Right image:  replicator dynamics stable states~\cite{weibull95,Hofbauer1998} 
with $50\%$ cooperators and defectors initially in a well mixed population for
comparison purposes.  Lighter tones stand for more cooperation. Values in parentheses next to each quadrant indicate average cooperation in the corresponding game space. 
 \label{PhaseSpace}
 }
\end{center}
\end{figure*}

There is an infinite number of games of each type since any positive affine transformation of the payoff matrix
leaves the NE set invariant~\cite{weibull95}. Here
we study the customary standard parameter space~\cite{santos-pach-06,anxo1}, by fixing the payoff values in the following
way: $R=1$, $P=0$, $-1 \leq S \leq 1$, and $0 \leq T \leq 2$. 
Therefore, in the $TS$ plane each game class corresponds
to a different quadrant depending on the above ordering of the payoffs as depicted in Fig.~\ref{PhaseSpace}, left image.
The right image depicts the well mixed replicator dynamics stable states for future comparison.

\subsection{Population Structure}
\label{pop-str}

The Euclidean two-dimensional space is modeled by a discrete square lattice of side $L$ with toroidal borders. 
Each vertex of the lattice can be occupied by one player or be empty.  The  \textit{density} is $\rho=N/L^2$, where 
 $N \le L^2$ is the number of players.
Players can interact with $k$ neighbours which lie at an Euclidean distance smaller or equal than a given constant $R_g$. 
Players can also migrate to empty grid points at a distance smaller than $R_m$. We use three neighborhood sizes
with radius $1.5$, $3$, and $5$; they contain, respectively, $8$, $28$, and $80$ neighbours around the central
player.




\subsection{Payoff Calculation and Strategy Update Rules}
\label{revision-protocols}

Each agent $i$ interacts locally with a set of neighbours $V_i$ lying closer than $R_g$.
\noindent Let $\sigma_i(t)$ be a vector
giving the strategy profile at time $t$ with $C= (1, 0)$ and $D = (0, 1)$ and let $M$ be the payoff matrix of the game (equation~\ref{eq:payoff}). 
The quantity
\begin{equation}
\Pi_i(t) =  \sum _{j \in V_i} \sigma_i(t)\; M\; \sigma_{j}^\top(t)
\label{payoffs}
\end{equation}
is the cumulated payoff collected by player $i$ at time step $t$.

We use two imitative strategy update protocols. The first is the Fermi rule in which 
the focal player $i$ is given the opportunity to imitate a randomly chosen neighbour $j$ with probability:
\begin{equation}
  p(\sigma_i \rightarrow \sigma_j)  = \frac{1}{ 1+exp(-\beta(\Pi_j - \Pi_i))}
\label{fermi}
\end{equation}
where $\Pi_j -\Pi_i$ is the difference of the payoffs earned by $j$ and $i$ respectively and
$\beta$ is a constant corresponding to the inverse temperature of the system. 
When $\beta \to 0$ the probability of imitating $j$ tends to a constant value $0.5$ and when $\beta \to \infty$ the rule becomes deterministic: $i$ imitates $j$ if  $(\Pi_j - \Pi_i)>0$, otherwise it doesn't. In between these two extreme cases the probability of imitating neighbour $j$ is an increasing function of $\Pi_j - \Pi_i$.

\noindent The second imitative strategy update protocol
is to switch to the strategy of the neighbour that has scored
best in the last time step. In contrast with the previous one, this rule is deterministic.
This \textit{imitation of the best} (IB) policy can be described in the following way:
the strategy $\sigma_i(t)$ of individual $i$ at time step $t$ will be
\begin{equation}
\sigma_i(t) = \sigma_j(t-1),
\label{ib}
\end{equation}
where
\begin{equation}
j \in \{V_i \cup i\} \;s.t.\; \Pi_j = \max \{\Pi_k(t-1)\}, \; \forall k \in \{V_i \cup i\}.
\label{ib2}
\end{equation}
\noindent That is, individual $i$ will adopt the strategy of the player with the highest
payoff among its neighbours including itself.
If there is a tie, the winner individual is chosen uniformly at random.

\subsection{Population Dynamics and Opportunistic Migration}
\label{migration}

We use an asynchronous scheme for strategy update and migration, i.e. players are updated one by one by choosing a random player in each step with uniform probability and with replacement. Then the player migrates with probability $1/2$, otherwise it updates its strategy. 
If the pseudo-random number drawn dictates that $i$ should migrate, then it considers $N_{test}$ randomly chosen positions in the disc of radius $R_m$ around itself. 
The quantity $N_{test}$ could be seen as a kind of ``energy'' available to a player for moving around and doing its search.
$N_{test}$ being fixed for a given run, it follows that an agent will be able to make a more complete exploration of its
local environment the smaller the $R_m$.
For each trial position the player computes the payoff that it would obtain in that place with its current strategy. The positions already occupied are just discarded from the possible choices. Then player $i$ stays at its current position if it obtains there the highest payoff, or migrates to the most profitable position among those explored during the test phase. 
 If several positions, including its current one, share the highest payoff then it chooses one at random. We call this migration
 \textit{opportunistic} or \textit{fitness-based}.
The protocol described in Helbing and Yu~\cite{helbingPNAS} is slightly different: the chosen player chooses the strategy
of the best neighbour including itself with probability $1-r$, and with probability $r$, with $r \ll 1-r$, its strategy is randomly reset. Before
this imitation step $i$ deterministically chooses the highest payoff free position in a square neighborhood of size $(2M+1) \times (2M+1)$ cells surrounding
the current player and including itself, where $M$ can take the values $0,1,2,5$. If several positions provide the same payoff, the one that is closer is selected. 
 
 \subsection{Simulation Parameters}
\label{Simulation Parameters} 

 The $TS$ plane has been sampled with a grid step of $0.1$ and
 each value in the phase space reported in the figures is the average of $50$ independent runs.  
 The evolution proceeds by first initializing the population by distributing $N=1000$ players with uniform probability among the
 available cells. Then  the players' strategies are initialized  uniformly at random
such that each strategy has a fraction of approximately $1/2$.  To avoid transient states, we let the system evolve for a period of $\tau=1000$  time steps and, in each time step, $N$ players are chosen for update.
 At this point almost always the system reaches a
 steady state in which the frequency of cooperators is stable except for small statistical fluctuations. We then let the system evolve for $50$ further
 steps and take the average cooperation value in this interval. We repeat the whole process $50$ times for each grid point
  and, finally, we report the average cooperation values over those $50$ repetitions.

\section{Results}

\subsection{Imitation of the Best and Opportunistic Migration}
\label{IBAM}

\begin{figure*}[ht!]
\begin{center}
\begin{tabular} {cccccccc} 
\includegraphics[width=6cm]{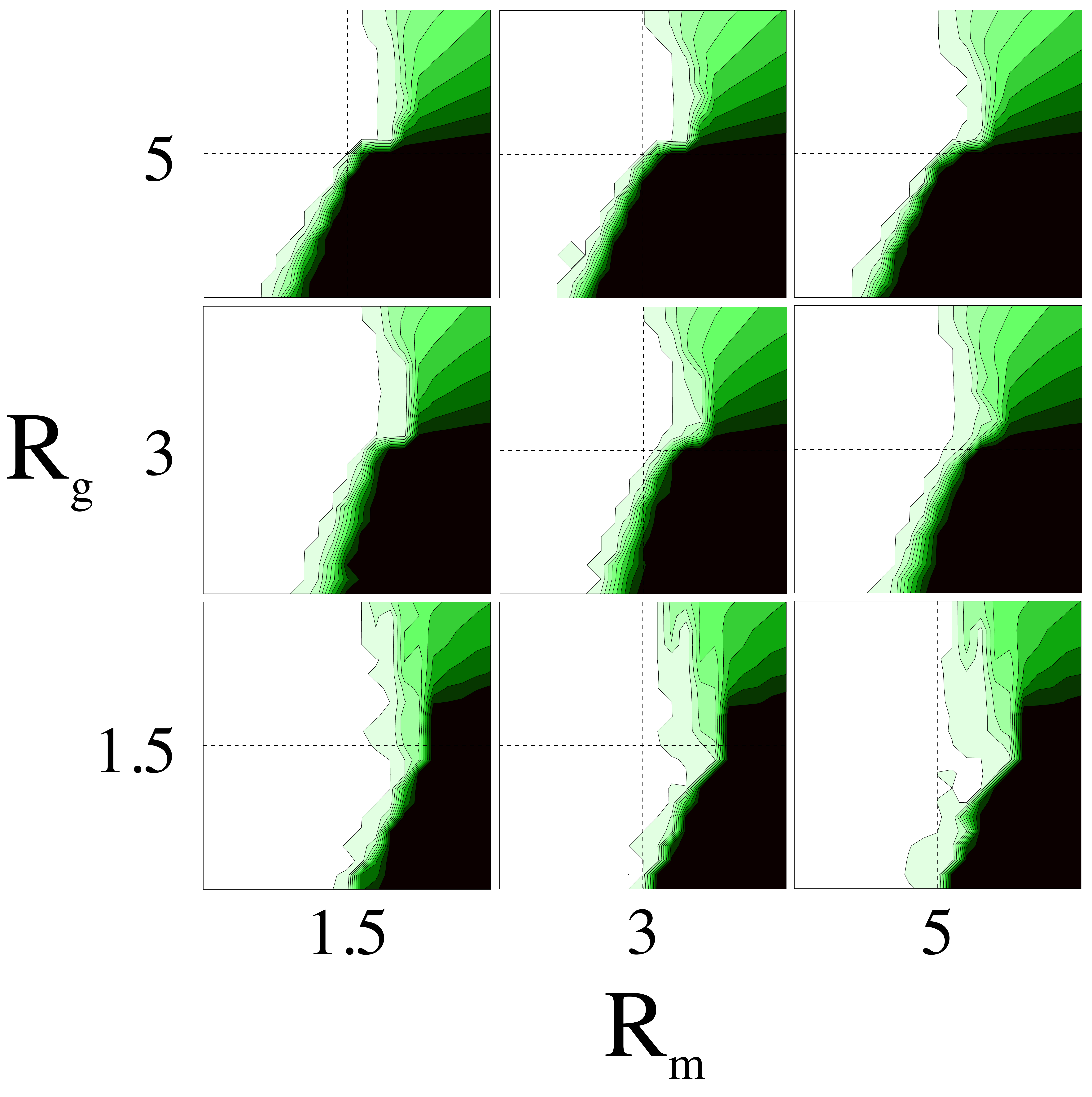} &
\includegraphics[width=6cm]{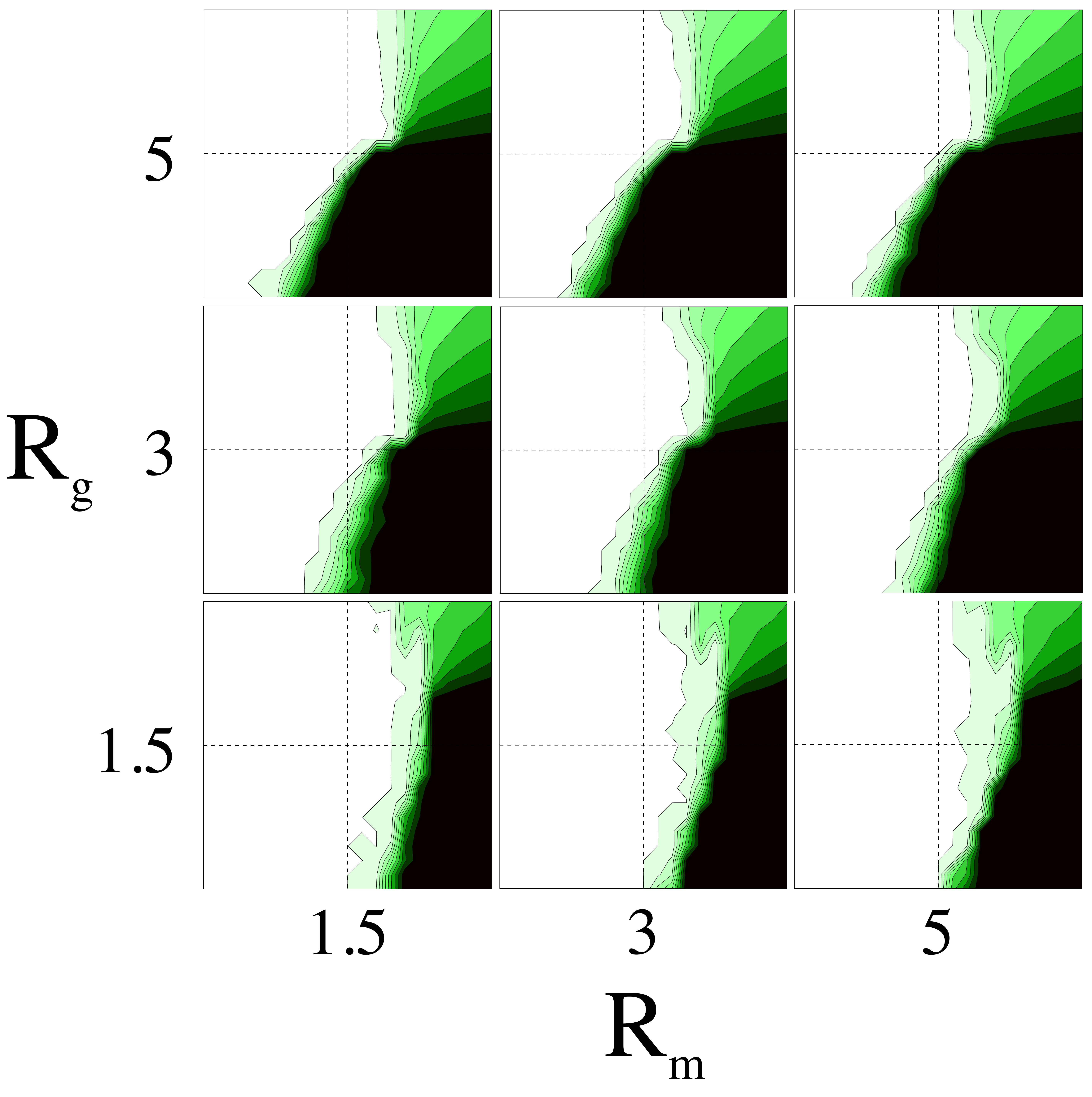} &
\end{tabular}
\caption{(Color online) Average cooperation levels with opportunistic migration and IB rule as a function of $R_g$ and $R_m$. Left: $N_{test}= 20$; Right: $N_{test} = 1$.  The size of the population is $1000$ players and the
density $\rho$ is $0.5$. In all cases the initial fraction of cooperators is $0.5$ randomly distributed among the occupied grid points. 
 \label{IB1}
 }
\end{center}
\end{figure*}

\begin{figure*}[ht!]
\begin{center}
\begin{tabular} {cccccccc} 
\includegraphics[width=7cm]{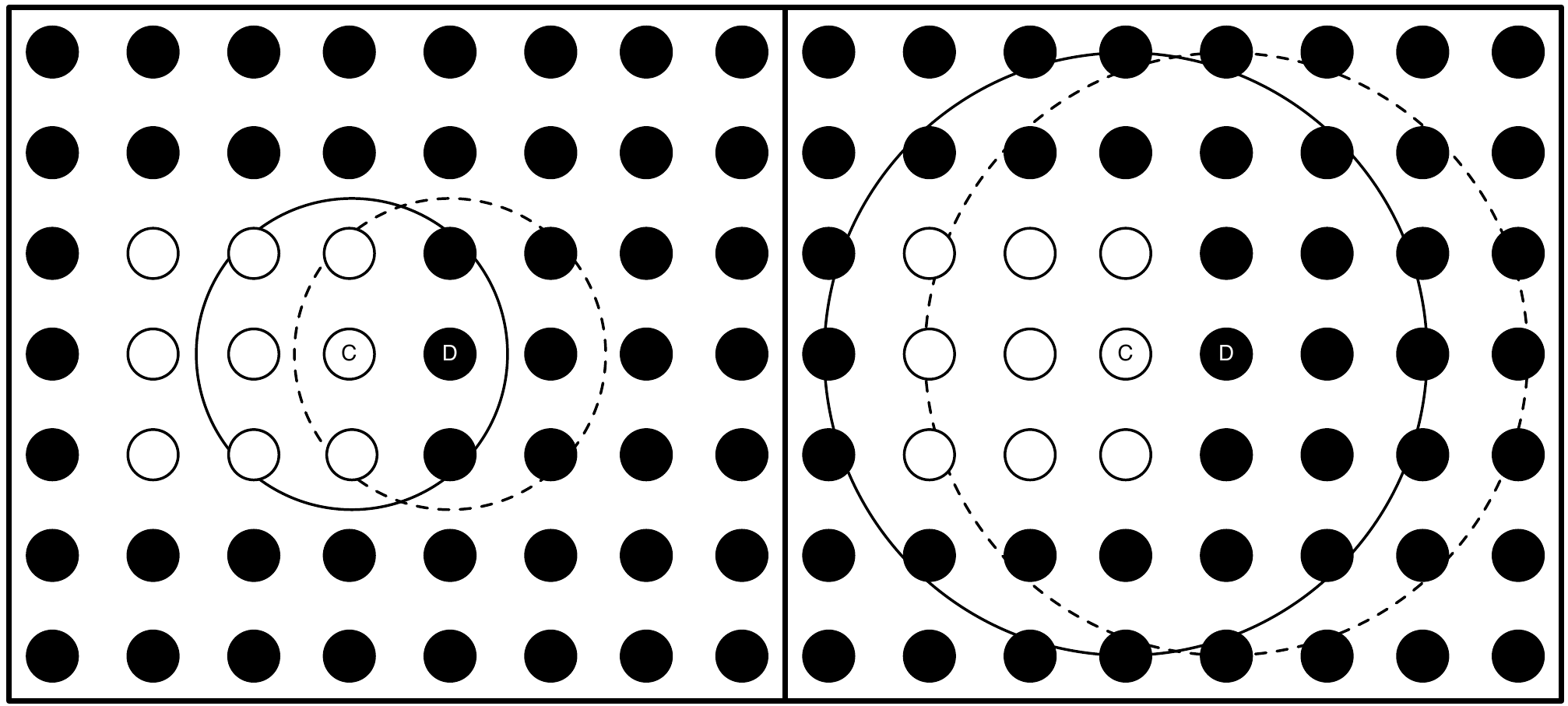} 
\end{tabular}
\caption{ Illustration of the effect of the playing radius $R_g$ on the payoff of individuals. Left image: $R_g=1.5$,
righ image: $R_g=3$. The drawings refer to a locally full grid and are intended for illustrative purposes only (see text). 
 \label{grids}
 }
\end{center}
\end{figure*}

In this section we study cooperation with the IB rule and fitness-based opportunistic migration, and we explore the influence of different radii $R_m$ and $R_g$ and other parameters such as the density $\rho$ and the number of trials $N_{test}$.
The left image of Fig.~\ref{IB1}  displays the TS plane with the IB rule, a density $\rho=0.5$, and $N_{test}= 20$. 
For small $R_g=1.5$ full cooperation is achieved in the SH quadrant for all $R_m$. The average levels of cooperation in the PD games are $0.33, 0.31,0.30$ for  $R_m = 1.5, 3, 5$ and $R_g = 1.5$ respectively.  It is remarkable that cooperation emerges
in contrast to the well mixed population case (Fig.~\ref{PhaseSpace}, right image), and also that better results are obtained with respect to a fully populated grid
 in which agents cannot move~\cite{anxo1}.
 The HD doesn't benefit in the same way and the cooperation levels are almost the same in the average.
 Cooperation remains nearly constant as a function of $R_m$ for a given $R_g$ value but increasing $R_g$
 has a negative effect.
 For higher game radius, $R_g\in\{3, 5\}$ cooperation is progressively lost in the PD games while there is little variation in the HD quadrant  among the different cases due to the dimorphic structure of these populations.
 In the SH quadrant
 there is a large improvement  compared to the well mixed case but the gain tends to decrease  with increasing $R_g$.  In the PD 
 with high $R_g$, cooperators cannot increase their payoff by clustering, since the neighborhood of defectors covers adjacent small clusters of cooperators, the payoff of defectors becomes higher and they can invade cooperators clusters. 
 Figure~\ref{grids} illustrates in an idealized manner what happens to a small cooperators cluster when the game radius $R_g$ increases using a full grid
 for simplicity. 
For $R_g=1.5$ (left image) the cooperator cluster is stable as long as $8R > 3T$ since the central
cooperator gets a payoff of $8R$, while the best payoff among the defectors is obtained by the individual marked $D$
(and by the symmetrically placed defectors) and is equal to $3T$ since $P=0$.  Under this condition all the
cooperators will thus imitate the central one. On the other hand, the defector will turn into a $C$ as long
as $5R+3S > 3T$, thus provoking cooperator cluster expansion for parameter values in this range. On the contrary,
for $R_g=3$ (right image) the central cooperator gets $8R+20S$ whilst the central defector at the border has
a payoff of $7T$. Thus the cooperator imitates the defector  if $7T > 8R + 20S$, i.e. $7T > 8 + 20S$ since 
$R=1$. This qualitative argument helps to explain the observed cooperation losses for increasing $R_g$.

This inequality is satisfied almost everywhere in the PD quadrant except in a very small area in its upper
left corner.

 Helbing and Yu~\cite{helbingPNAS} found very encouraging cooperation results in their analysis but they only had a small
 game radius corresponding to the Von Neumann neighborhood which is constituted, in a full lattice, by the central
 individual and the four neighbours at distance one situated north, east, south, and west. We also find similar results for
our smallest neighborhood having $R_g=1.5$, which corresponds to the eight-points Moore neighborhood but, as $R_g$ gets larger, we have just seen
 that a sizable portion of the cooperation gains are lost. We think that this is an important point since there are certainly
 situations in which those more extended neighborhoods are the natural choice in a spatially extended population.

  \begin{figure*}[ht!]
\begin{center}
\begin{tabular} {cccccccc} 
\includegraphics[width=6cm]{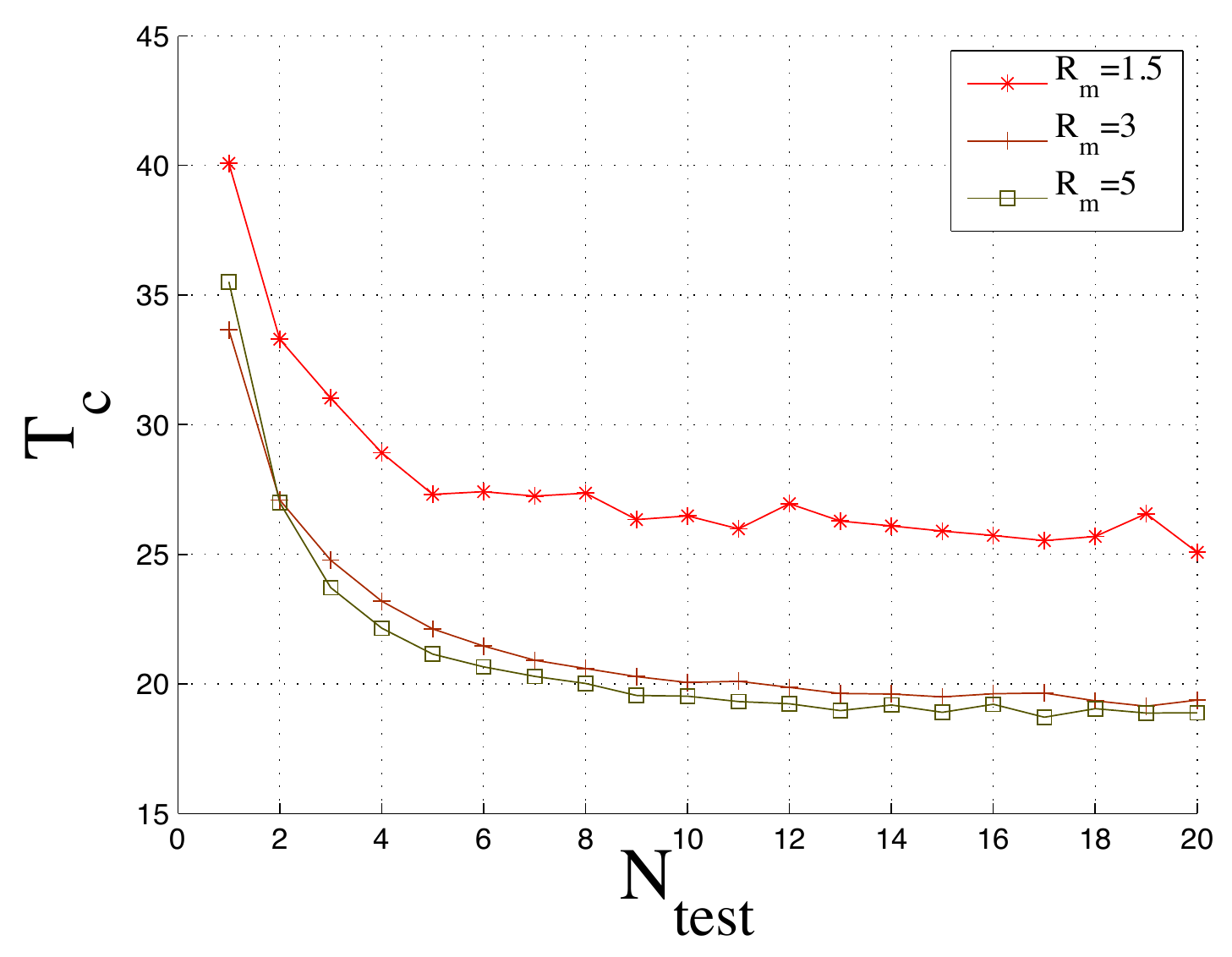} &
\includegraphics[width=6cm]{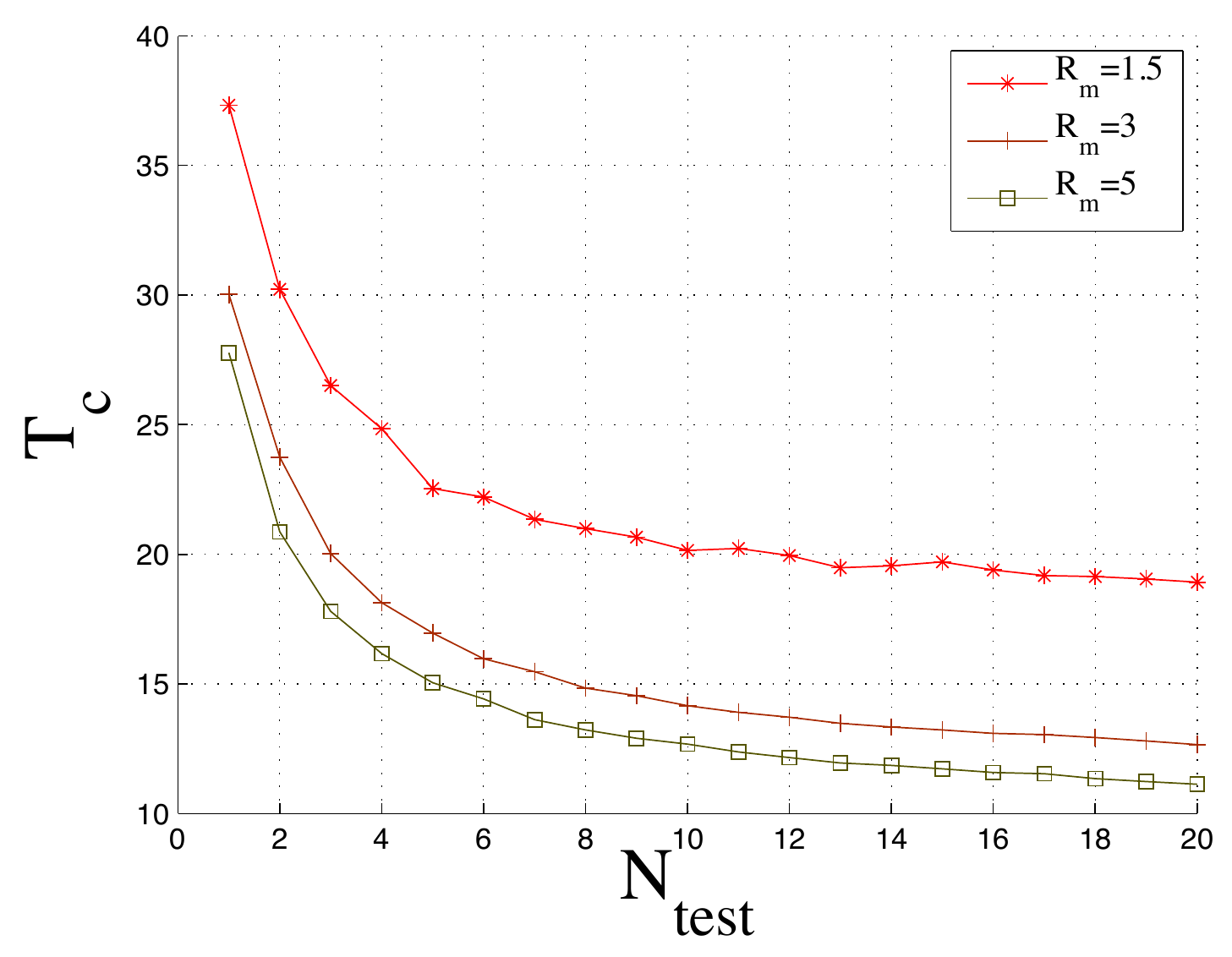} &
\end{tabular}
\caption{(Color online)  Average convergence time $T_c$ with IB rule as a function of $N_{test}$ for $R_g=1.5$ and $R_m=1.5,3,5$. Left image: $S=-0.5$, $T=0.5$. Right image: $S=-0.1$, $T=1.1$. The time to convergence is defined as the number of 
simulation steps needed for the number of cooperators $N_c$ or defectors $N_d$ to be smaller than $0.1N$. 
Times of convergence are averaged over $500$ independent runs.
 \label{ConvergenceTimes}
 }
\end{center}
\end{figure*}

The number of trials $N_{test}$ could also be a critical parameter in the model. The right image of Fig.~\ref{IB1} refers to the same case as the left one, 
i.e. the IB update rule with opportunistic migration and $\rho=0.5$, except for the number of trials which is one instead
of $20$. We observe that practically the same cooperation levels are reached at steady state in both cases for $R_m = 5$ and 
$R_m=3$, while there is a small increase of the average cooperation in the PD games for $R_g = 1.5$ which
goes from $0.33$, $0.31$, and $0.30$ for $N_{test}=20$ to $0.41$, $0.36$, and $0.33$ for $N_{test}=1$,  for $R_m=1.5,3,5$ respectively. 
On the whole, it is apparent that $N_{test}$ does not seem to have a strong influence.
However, one might
ask whether the times to convergence are shorter when more tests are used, a fact that could compensate for the extra work
spent in searching. But Figs.~\ref{ConvergenceTimes}
show that convergence times are not very different and decrease very quickly with the number of essays $N_{test}$. This
is shown for two particular games, one in the middle of the SH quadrant (left image), and the other near the
upper left corner of the PD space (right image). Thus, a shorter time does not compensate for the wasted trials. 
Since moving around to find a better place is a costly activity in any real situation, this result is encouraging because it says that 
searching more intensively doesn't change the time to convergence for more than four tests. 
 Thus, quite high levels of cooperation can be achieved by opportunistic migration
at low search cost, a conclusion that interestingly extends the results presented in~\cite{helbingPNAS}.

 
In diluted grids, density is another parameter that influences the evolution of cooperation~\cite{Arenzon1,Arenzon2}, also in the
presence of intelligent migration~\cite{helbingPNAS,adaptive-mig}. Too high densities  should be detrimental
because clusters of cooperators are surrounded by a dense population of defectors, while low densities allow cooperator clusters to have less defectors in their neighborhood once they are formed. 
We have performed numerical simulations for two other values of the density besides $0.5$, $\rho=0.2$ and $\rho=0.8$.
We do not show the figures to save space but the main remark is that there is a monotone decrease of cooperation going from low to higher densities in the low
$S$ region that influences mainly the PD and, to a smaller extent, the SH games.

\subsection{Opportunistic Migration and Noisy Imitation}

\begin{figure*}[ht!]
\begin{center}
\begin{tabular} {cccccccc} 
\includegraphics[width=3.7cm]{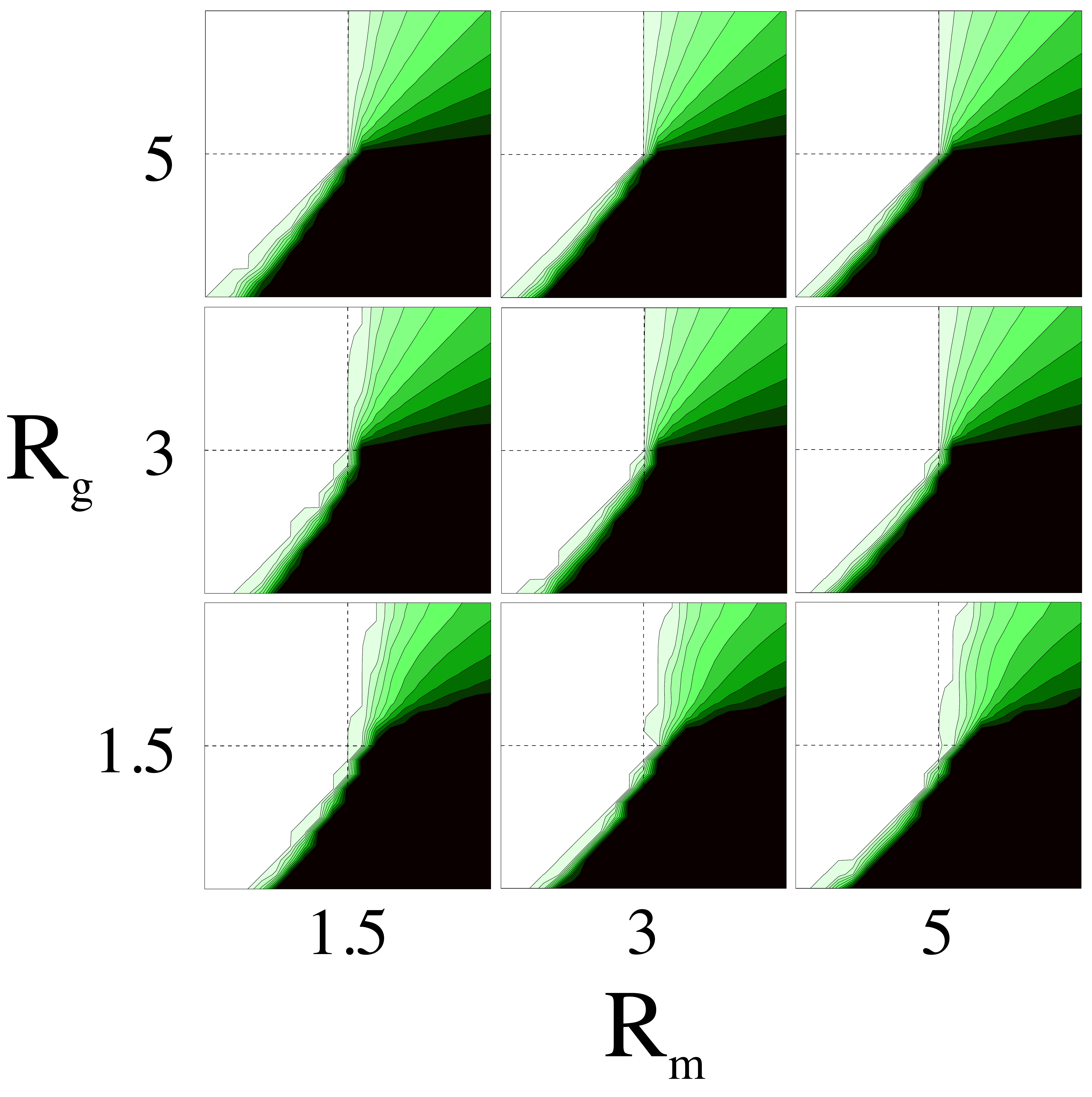} &
\includegraphics[width=3.7cm]{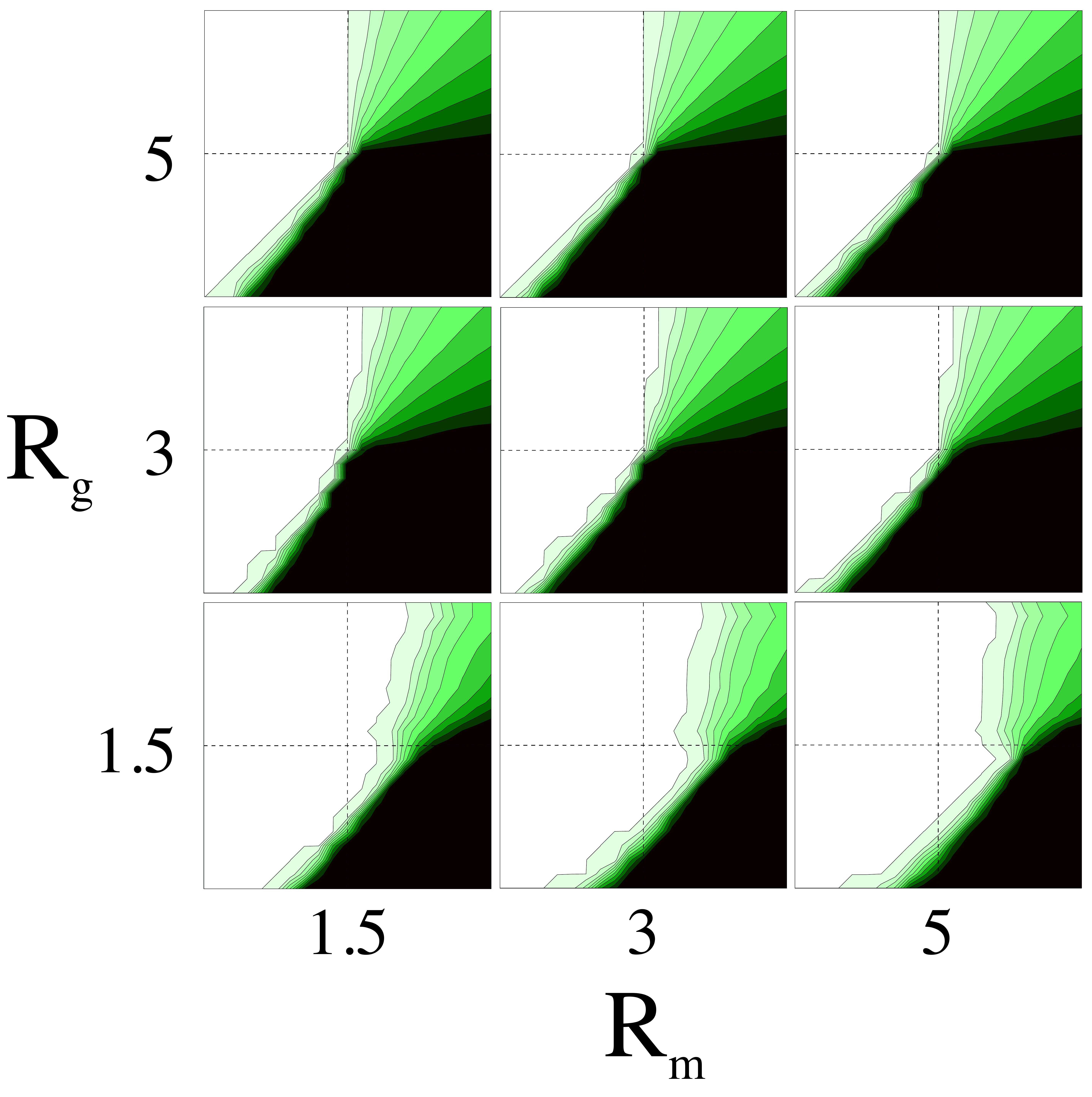} &
 \includegraphics[width=3.7cm]{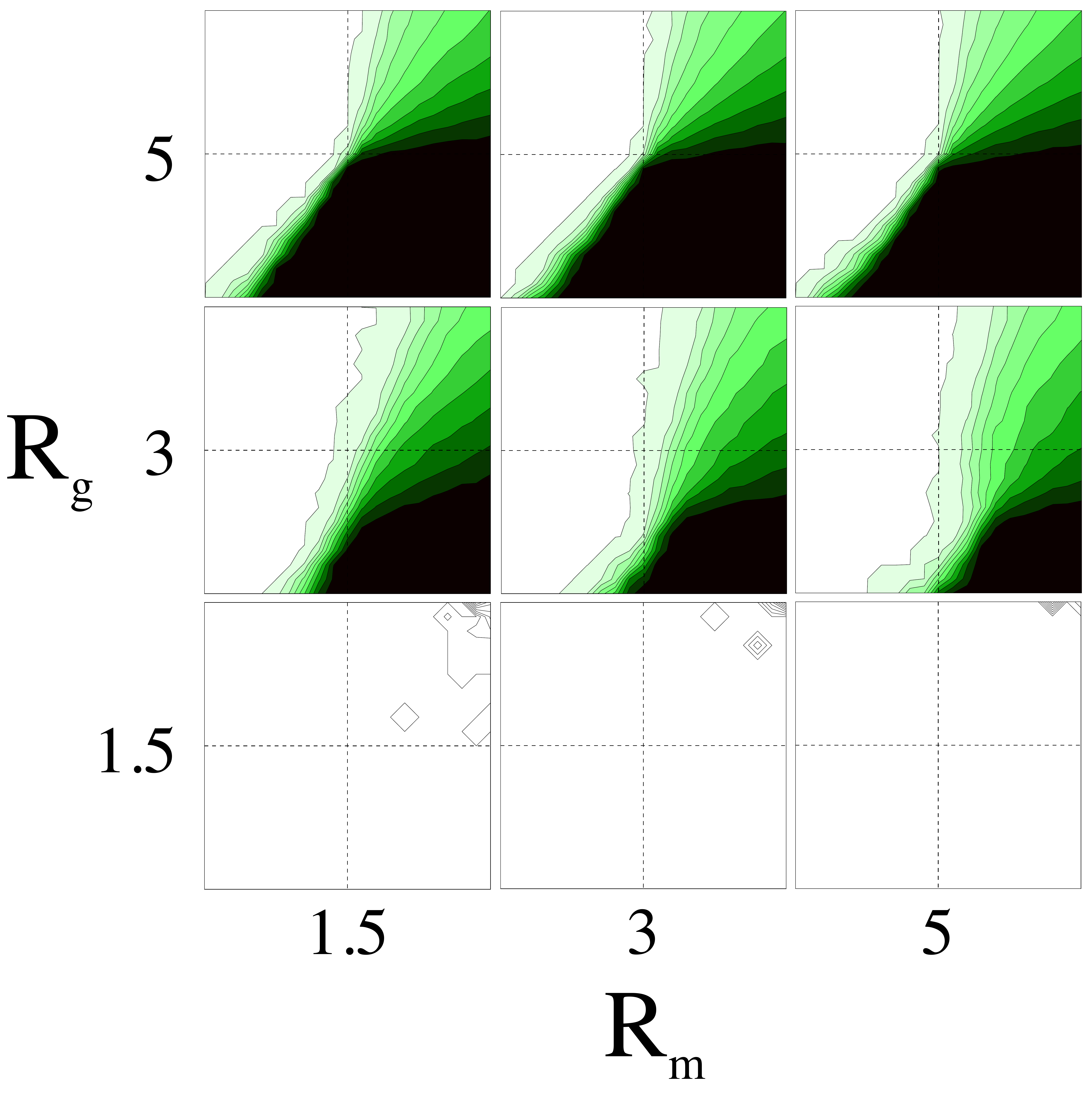} &
  \includegraphics[width=3.7cm]{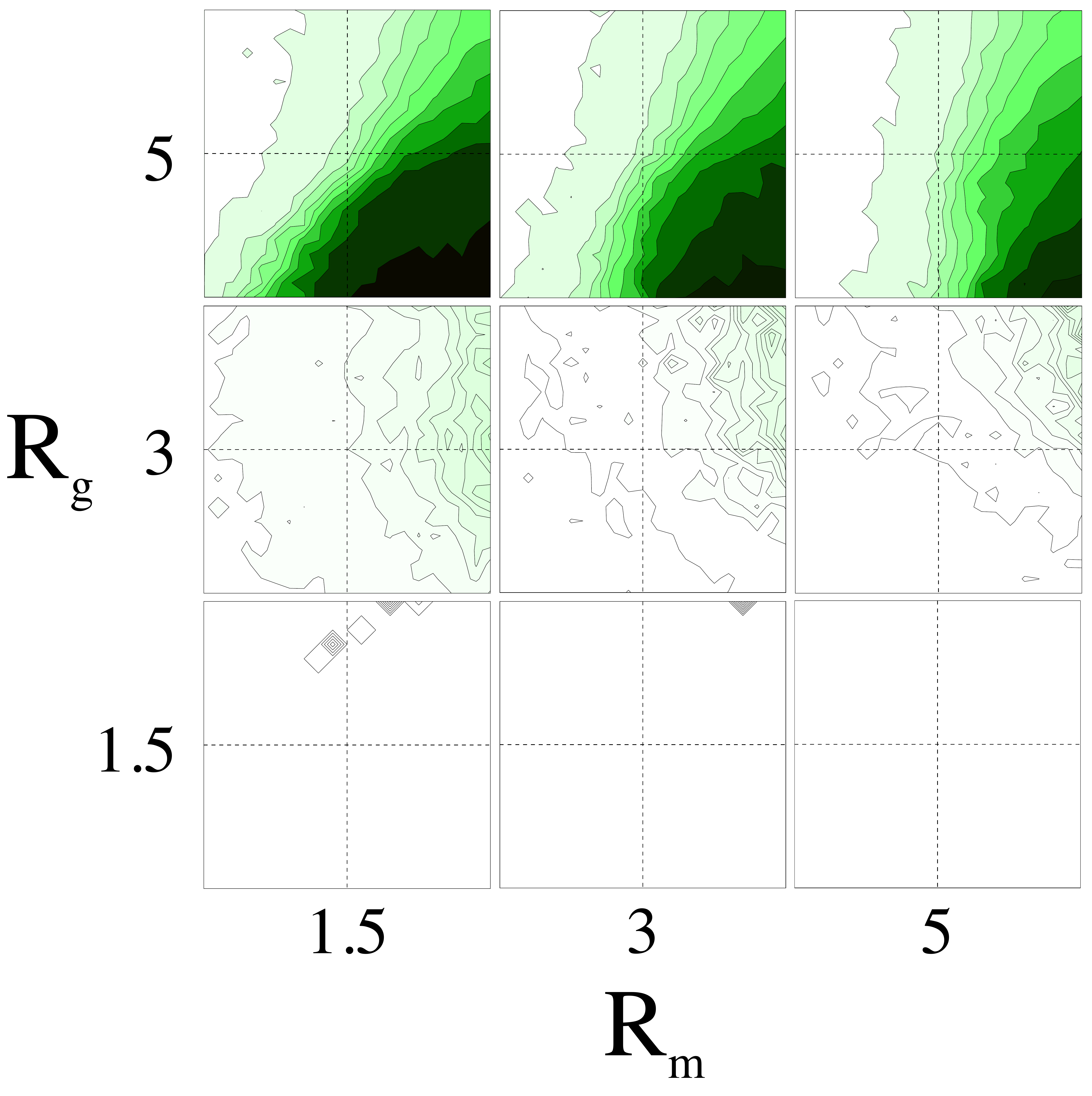} &
\end{tabular}
\caption{(Color online)  Average cooperation levels with opportunistic migration  and the Fermi rule as a function of $R_g$ and $R_m$. From left to right $\beta = 1.0, 0.1, 0.01, 0.001$. The density $\rho$ is $0.5$ and $N_{test}= 20$. The size of the population is $1000$ players. In all cases the initial fraction of cooperators is $0.5$ randomly distributed among the population.
 \label{Fermi0p5}
 }
\end{center}
\end{figure*}

In this section we use the more flexible strategy update protocol called the Fermi rule which was  described in Sect.~\ref{revision-protocols} and in which the probability to imitate a random neighbour's strategy depends on the parameter $\beta$. 
We have seen that using the IB rule with adaptive migration leads to full cooperation in the SH quadrant and improves cooperation 
in a part of the PD quadrant (Fig.~\ref{IB1}). This result does not hold  with the Fermi rule with $\beta\ge1$, and we are back to
full defection in the PD and almost $50\%$ cooperation as in the well mixed case in the SH; this behavior can be
appreciated in the leftmost image of Fig.~\ref{Fermi0p5}. 

\begin{figure*}[ht!]
\begin{center}
\includegraphics[width=14cm]{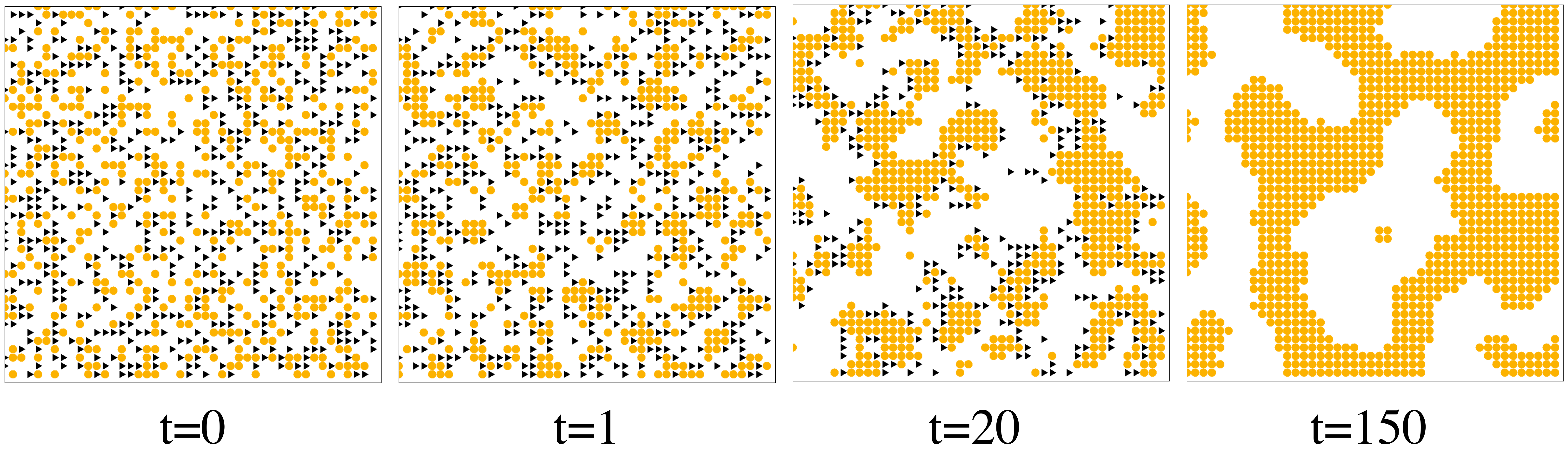} 
\caption{(Color online). Time evolution for the case of a PD with $T=1.5$, $S=-0.5$, $R=1$, $P=0$. Here
$\beta=0.01$, $R_g=1.5$, $R_m=3$. The density $\rho=0.5$ and $N_{test}=20$. There are $1000$ players and the
strategies are initially attributed uniformly at random in a $50-50$ proportion.
 \label{FermiRand}
 }
\end{center}
\end{figure*}

An interesting new phenomenon appears when $\beta$ becomes small, of the order of $10^{-2}$. In this case, 
the levels of cooperation increase in all games  for $R_g$ values up to $3$ and cooperation raises to almost $100\%$ in 
all game phase space for
$R_g=1.5$, for all migration radii, see the third image of Fig.~\ref{Fermi0p5}. The positive trend continues with decreasing
$\beta$ (see rightmost image) and cooperation prevails almost everywhere.
As we said above, the Fermi rule with $\beta=0.01$ or less implies
that the decision to imitate a random neighbour becomes almost random itself. Thus, the spectacular gains in cooperation
must depend in some way from opportunistic migration for the most part. Figure~\ref{FermiRand} illustrates the
dynamical behavior of a particular case in the PD space. Here $T=1.5$, $S=-0.5$, $R=1$, $P=0$; that is, the game is in the
middle of the PD quadrant. The other parameters are: $\beta=0.01$,
$R_g=1.5$, and $R_m=3$. This particular game would lead to full defection in almost all cases but here we can see
that it leads to full cooperation instead.

This is a surprising phenomenon that needs an explanation. At the beginning, due to opportunistic migration,
cooperators will be likely to form small clusters between themselves more than defectors, as the latter tend to follow
cooperators instead of clustering between themselves since the $(D,D)$ payoff is equal to $0$. 
The low $\beta$ value will make strategy change close
to random and thus strategy update will have a neutral effect. 
Indeed, as soon as cooperator clusters form due
to migration, defectors that enter a cooperator cluster thanks to random imitation cannot invade them.
The situation there is akin to a full grid and the number of defectors inside the cluster will fluctuate. Meanwhile, defectors at the
border of a cooperator cluster will steadily turn into cooperators thus extending the cluster. This is due to the fact
that lone defectors at the border will tend to imitate cooperators since
defectors are less connected, and strategy imitation is almost random. 
Finally, the defectors inside the clusters
will reach the border and turn into cooperators as well. The phenomenon is robust with respect to the migration
radius $R_m$, as can be seen in the lower part of the third and fourth images of Fig.~\ref{Fermi0p5}. Cooperation prevails
even when $P$ becomes positive which increases the payoff for defectors to aggregate. We have simulated
the whole phase space for $P=0.2$ and $P=-0.2$. The results are similar to those with $P=0$ except that cooperation decreases slightly with increasing $P$.
On the same images it can be seen that the game radius $R_g$ has a large influence and cooperation
tends to be lost for radii larger than $1.5$. The reasons for this are very similar to those advocated in Sect.~\ref{IBAM}
where Fig.~\ref{grids} schematically illustrates the fact that increasing $R_g$ makes the situation more similar
to a well mixed population. In these conditions, the payoff-driven strategy imitation process becomes more important and
may counter the benefits of opportunistic migration. However, since we believe that system possessing locality 
are important in practice, the findings of this section seem very encouraging for mobile agents that are better
at finding more profitable positions and moving to them rather than at strategic reasoning.

\section{Discussion and Conclusions}

In this work we have explored some possibilities that arise when agents playing simple two-person, two-strategy
evolutionary games may also move around in a certain region seeking better positions for themselves. The games
examined are the standard ones, like the Prisoner's Dilemma, the Hawk-Dove, or the Stag Hunt. 
In this context, the ability to move around in space is 
extremely common in animal as well as human societies and therefore its effect on global population behavior is
an interesting research question. As already pointed out by other 
researchers~\cite{helbingPNAS,Helb-Mobil,adaptive-mig,reput-migr,SwarmPD,aktipis}, adding a form of contingent mobility may
result in better capabilities for the population to reach socially valuable results. Among the existing models, we have started
from a slightly modified form of the interesting Helbing's and Yu's model~\cite{helbingPNAS} and 
have explored some further avenues that were left untouched in the latter work.
In the model agents live and move in a discrete two-dimensional grid space in which part of the cells are unoccupied.
Using a strategy update rule that leads an agent to imitate her most successful neighbour as in~\cite{helbingPNAS}, and
having the possibility to explore a certain number of free positions around oneself to find a better one, the gains in 
cooperative behavior are appreciable in the Prisoner's Dilemma, in qualitative agreement with~\cite{helbingPNAS}.
In the Hawk-Dove games the gains in cooperation are small but, in addition, we find that cooperation is fully promoted 
in the class of Stag Hunt games which were not considered in~\cite{helbingPNAS}.
In Helbing and Yu the exploration of the available cells in search of a better one was fixed and
deterministic. The question of the amount of effort needed to improve the agent's situation was left therefore open,
although this is clearly an important point, given that in the real world more exploration usually entails an increasing
cost be it in terms of energy, time, or money. By using a similar search strategy but to random positions within a given radius, and by varying  the number of searches available to
the agent, we have seen that the convergence times to reach a given average level of cooperation do not degrade 
significantly by using fewer trials.
This is a reassuring finding, given the above remarks related to the search cost. 

Helbing and Yu explored migration effects under a number of sizes of the square neighborhood around a given agent.
However, they only had a single neighborhood for the game interactions, the standard five-cells Von Neumann neighborhood.
We have explored this aspect more deeply and presented results for several combinations of game radius $R_g$ 
and migration radius $R_m$. In fact, it turns out that increasing the interaction radius has an adverse effect on cooperation
to the point that, at $R_g=5$, cooperation levels are similar to those of a well mixed population, in 
spite of fitness-based migration. Thus, positive results are only obtained when agents interact locally in a relatively small
neighborhood which, fortunately, seems to be a quite common condition in actual spatial systems.


Most importantly, we have explored another important commonly used strategy update rule, the Fermi rule. This rule
is also imitative but allows to control the intensity of selection by varying a single parameter $\beta$. When
$\beta$ is high, i.e. larger than one, almost all the cooperation gains observed with the imitation of the best rule are
lost and we are back to a scenario of defection in the Prisoner's Dilemma space and the Stag Hunt games
are also influenced negatively. Migration does not help in this case. However, when $\beta$ is low, of the order of $0.01$, a very interesting phenomenon
emerges: cooperation prevails everywhere in the game space for small game radius and for all migration radii,
including in the PD space, which is notoriously the most problematic class of games.
With $\beta=0.01$ or lower the strategy update is close to random; however, fitness-based migration is active and 
thus we see  that migration, and not strategy update, is the main force driving the population towards cooperation
and we have hypothesized a qualitative mechanism that could explain this striking result.
Cooperation is robust with respect to the migration radius $R_m$ but increasing $R_g$ affects the results negatively
for $R_g \ge 3$. The effect is mitigated the more random the strategy update, i.e. by further decreasing
$\beta$.

\subsection{Acknowledgments} The authors thank the Swiss National Foundation for their financial support under contracts  200021-14661611 and 200020-143224.



\end{document}